\documentclass[aps,prc,twocolumn,superscriptaddress,showpacs]{revtex4}
\usepackage{graphicx}
\usepackage{dcolumn}

\begin{document}

\title{Multi-quasiparticle structures up to spin $\sim{44}\hbar$ in the odd-odd nucleus $^{168}$Ta}

\author{X. Wang}
\affiliation{Department of Physics, Florida State University, Tallahassee, FL 32306}
\author{D. J. Hartley}
\affiliation{Department of Physics, United States Naval Academy, Annapolis, MD 21402}
\author{M. A. Riley}
\affiliation{Department of Physics, Florida State University, Tallahassee, FL 32306}
\author{L. L. Riedinger}
\affiliation{Department of Physics and Astronomy, University of Tennessee, Knoxville, TN 37996}
\author{A. Aguilar}
\affiliation{Department of Physics, Florida State University, Tallahassee, FL 32306}
\author{M. P. Carpenter}
\affiliation{Physics Division, Argonne National Laboratory, Argonne, IL 60439}
\author{C. J. Chiara}
\altaffiliation{Present address: Department of Chemistry and Biochemistry, University of Maryland, College Park, MD 20742}
\altaffiliation{Physics Division, Argonne National Laboratory, Argonne, IL 60439}
\affiliation{Nuclear Engineering Division, Argonne National Laboratory, Argonne, IL 60439}
\author{P. Chowdhury}
\affiliation{Department of Physics, University of Massachusetts Lowell, Lowell, MA 01854}
\author{I. Darby}
\affiliation{Department of Physics and Astronomy, University of Tennessee, Knoxville, TN 37996}
\author{U. Garg}
\affiliation{Physics Department, University of Notre Dame, Notre Dame, IN 46556}
\author{Q. Ijaz}
\affiliation{Department of Physics, Mississippi State University, Mississippi State, MS 39762}
\author{R. V. F. Janssens}
\affiliation{Physics Division, Argonne National Laboratory, Argonne, IL 60439}
\author{F. G. Kondev}
\affiliation{Nuclear Engineering Division, Argonne National Laboratory, Argonne, IL 60439}
\author{S. Lakshmi}
\affiliation{Department of Physics, University of Massachusetts Lowell, Lowell, MA 01854}
\author{T. Lauritsen}
\affiliation{Physics Division, Argonne National Laboratory, Argonne, IL 60439}
\author{W. C. Ma}
\affiliation{Department of Physics, Mississippi State University, Mississippi State, MS 39762}
\author{E. A. McCutchan}
\affiliation{Physics Division, Argonne National Laboratory, Argonne, IL 60439}
\author{S. Mukhopadhyay}
\affiliation{Department of Physics, Mississippi State University, Mississippi State, MS 39762}
\author{E. P. Seyfried}
\affiliation{Department of Physics, United States Naval Academy, Annapolis, MD 21402}
\author{I. Stefanescu}
\affiliation{Department of Chemistry and Biochemistry, University of Maryland, College Park, MD 20742}
\affiliation{Physics Division, Argonne National Laboratory, Argonne, IL 60439}
\author{S. K. Tandel}
\author{U. S. Tandel}
\affiliation{Department of Physics, University of Massachusetts Lowell, Lowell, MA 01854}
\author{C. Teal}
\altaffiliation{Present address: Nuclear Regulatory Commission, T4L7 11555 Rockville Pike, Rockville, MD 20855. 
Disclaimer: The views presented in this paper represent those of the authors alone, and not necessarily 
those of the Nuclear Regulatory Commission.}
\affiliation{Department of Physics, Florida State University, Tallahassee, FL 32306}
\author{J. R. Vanhoy}
\affiliation{Department of Physics, United States Naval Academy, Annapolis, MD 21402}
\author{S. Zhu}
\affiliation{Physics Division, Argonne National Laboratory, Argonne, IL 60439}

\date{\today}

\begin{abstract}
High-spin states in the odd-odd nucleus $^{168}$Ta have been populated in the $^{120}$Sn($^{51}$V,3n) reaction. 
Two multi-quasiparticle structures have been extended significantly from spin $\sim{20\hbar}$ to above 
${40\hbar}$. As a result, the first rotational alignment has been fully delineated and a second 
band crossing has been observed for the first time in this nucleus. 
Configurations for these strongly-coupled rotational bands are proposed based on signature 
splitting, $B(M1)/B(E2)$ ratio information, and observed rotation-alignment behavior. 
Properties of the observed bands in $^{168}$Ta are compared to related structures in the neighboring 
odd-$Z$, odd-$N$, and odd-odd nuclei and are discussed within the framework of the cranked shell model. 
\end{abstract}


\maketitle

\section{\label{sec:intro}Introduction}

Collective structures at high spin in the rare-earth nuclei ($\it{e.g.}$, $_{68}$Er, $_{69}$Tm, $_{70}$Yb, $_{71}$Lu, 
$_{72}$Hf, $_{73}$Ta) of the mass $\sim$ 160--170 region have become a focus of recent 
studies, since a number of these are associated with rather exotic excitation 
modes~\cite{Odegard-PRL-86-5866-01,Djong-PLB-560-24-03,Hartley-PRC-72-064325-05,Paul-PRL-98-012501-07,Pattabi-PLB-163Tm-07,Wang-PRC-163Tm-07,Aguilar-PRC-77-R021302-08}. 
More specifically, wobbling bands~\cite{Bohr-75-book}, characteristic of triaxial deformation, have drawn much attention. 
The latter have been interpreted as structures built on the $i_{13/2}$ proton orbital. However, 
it remains puzzling that this excitation mode has only been confirmed experimentally thus far in the odd-mass Lu 
isotopes~\cite{Bringel-EPJA-24-167-05,Odegard-PRL-86-5866-01,Jensen-PRL-89-142503-02,Schonwa-PLB-552-9-03,Amro-PLB-553-197-03} 
and in $^{167}$Ta~\cite{Hartley-PRC-80-R041304-09}. In fact, evidence for wobbling in $^{167}$Ta was found as 
part of a systematic search for this phenomenon in Ta nuclei (having two more protons than 
Lu)~\cite{Hartley-PRC-72-064325-05,Hartley-PRC-74-054314-06,Aguilar-170Ta-unpublish-10}. 
Owing to the power of Gammasphere, a considerable amount of data on $^{168}$Ta was also collected as a 
byproduct of this search. Although these data do not provide evidence for wobbling in $^{168}$Ta, they 
enabled new insight in the high-spin structure of this nucleus about which little was known. 

Compared to even-even and odd-mass nuclei, the properties of odd-odd systems are more challenging 
to study due to the complexity of level structures associated with contributions from both valence 
protons and neutrons. Consequently, only limited information is usually available for odd-odd nuclei throughout 
the chart of nuclides, even though these systems can aid significantly in our understanding of the 
dependence of band crossing frequencies and aligned angular momenta on the occupation of specific orbitals. 
The light odd-odd Ta nuclei are characterized by fairly small quadrupole deformations 
and provide a good opportunity to investigate this issue~\cite{Theine-NPA-536-418-92}. 
A recent high-spin study on $^{170}$Ta~\cite{Aguilar-170Ta-unpublish-10} by the present collaboration 
revealed 11 bands and allowed for a detailed investigation of these crossings. The data on $^{168}$Ta 
discussed here can be viewed as an extension of this work. 

Thus far, only two bands with spins up to $\sim{20}\hbar$ are known in $^{168}$Ta~\cite{Theine-NPA-536-418-92}. 
The purpose of the present work was three-fold: first, to confirm the previously 
known states at low spin and verify their properties (configuration, spin, and parity); 
second, to expand the level scheme by searching for new levels and structures; and third, to obtain a 
satisfactory interpretation of the experimental observations. 

\section{\label{sec:experiment}Experimental Details}
High-spin states in $^{168}$Ta were populated with a 235-MeV $^{51}$V beam delivered by 
the ATLAS facility at Argonne National Laboratory. The beam bombarded a stacked target of two 
self-supporting $^{120}$Sn foils, each with a thickness of 500 ${\mu}{\rm g}/{\rm cm}^{2}$. The $\gamma$ rays emitted 
in the reaction were collected by the Gammasphere array~\cite{Janssens-NPN-6-9-96} which consisted of 101 Compton-suppressed 
Ge detectors. Approximately $2{\times}10^{9}$ coincidence events with fold 4 and higher were 
accumulated during five days of beam time. Although the 4n (leading to $^{167}$Ta) and 5n ($^{166}$Ta) reaction 
channels dominated, a useful amount of data on $^{168}$Ta ($3n$-channel) were also acquired. The off-line data 
analysis was performed with the Radware software package~\cite{Radford-NIMA-361-297-95}. Hence, 
cubes and hypercubes were generated to establish the $\gamma$-ray coincidence relationships. 

\section{\label{sec:results}Results}

The states at low spin in $^{168}$Ta were first observed by Meissner $\it{et~al.}$~\cite{Meissner-ZPA-337-45-90} 
in the ${\beta}^{+}$ decay of $^{168}$W. In subsequent work by Theine $\it{et~al.}$~\cite{Theine-NPA-536-418-92}, 
two coupled $\Delta{I}=1$ rotational bands with spins up to $22\hbar$ and $20\hbar$ were identified and 
configurations were proposed based upon the fact that (i) the yrast band in the odd-$Z$ neighbor $^{167}$Ta 
(also studied in Ref.~\cite{Theine-NPA-536-418-92}) is built on a ${h}_{11/2}$ proton 
and that (ii) a $\nu{i}_{13/2}$ structure is yrast in $^{167}$Hf (odd-$N$ neighbor)~\cite{Janssens-PLB-106-475-81}. 
In $^{167}$Ta, a strong $\pi{d}_{5/2}$ band has also been observed. The concept of the additive coupling of quasiproton 
and quasineutron states in neighboring odd-mass nuclei~\cite{Kreiner-PRC-36-2309-87} was employed to propose 
tentative spin and parity assignments to the bands observed in $^{168}$Ta~\cite{Theine-NPA-536-418-92}. 

It is worth pointing out that, in the previous high-spin work of Ref.~\cite{Theine-NPA-536-418-92}, the 
transitions belonging to $^{168}$Ta were unambiguously identified by combining the following 
criteria: (i) all strong transitions were found to be in coincidence with Ta characteristic $x$ rays; 
(ii) many of the strong transitions were also found to be in coincidence with $A=168$ evaporation residues 
identified in another experiment where a recoil mass separator was used~\cite{Mo-private-commu}; 
and, (iii) all transitions at high spin were in coincidence with known 
low-lying transitions. Since the main purpose in the present work was to search for weakly 
populated states and bands at high spin, the expansion of level structures in $^{168}$Ta 
was based solely on coincidence relationships between newly observed $\gamma$ rays and known transitions 
of Ref.~\cite{Theine-NPA-536-418-92}. 

\begin{figure}
\begin{center}
\includegraphics[angle=0,width=0.8\columnwidth]{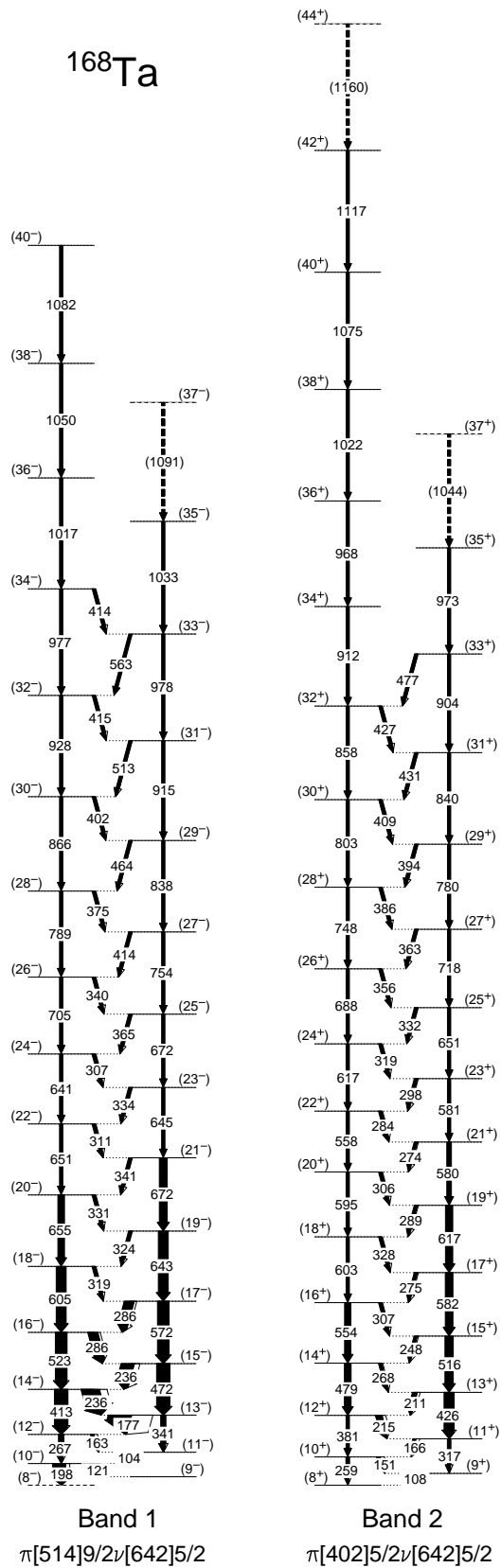}
\caption{Level scheme of $^{168}$Ta resulting from the present work. All spin and parity assignments are 
tentative (labels are put in parentheses) and are based on the proposed configurations. The states and 
transitions drawn as dashed lines are tentative. See text for details.\label{fig:168Ta_level_scheme}}
\end{center}
\end{figure}

\begin{figure*}
\begin{center}
\includegraphics[angle=270,width=1.6\columnwidth]{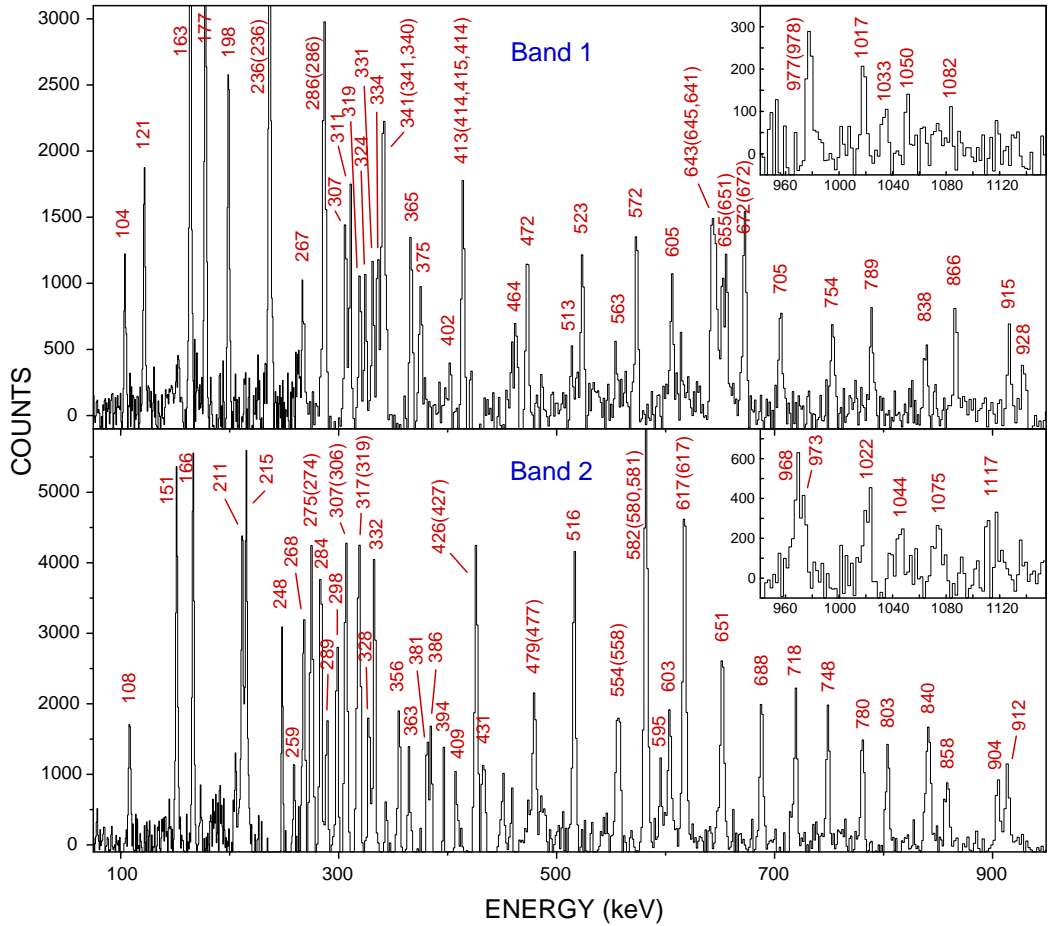}
\caption{(Color on-line) Coincidence spectra representative of bands 1 (top panel) 
and 2 (bottom panel) in $^{168}$Ta. These summed spectra were obtained by triple gating on selected 
in-band transitions, $267{\rightarrow}1017$ and $472{\rightarrow}1033$ of band 1 
and $381{\rightarrow}1117$ and $317{\rightarrow}973$ of band 2 (in keV). 
The highest-energy $\gamma$ rays observed in these bands are given in the insets. 
Transitions are labeled by their energies in keV. The labels with the ``number(number)'' format denote 
$\gamma$-ray multiplets, such as ``479(477)'' pointing to the 479-keV transition mixed with 
the 477-keV line, for example.\label{fig:168Ta_band1_2_spe}}
\end{center}
\end{figure*}

In the level scheme proposed in Fig.~\ref{fig:168Ta_level_scheme}, many $\gamma$ rays are degenerate 
or very close in energy to others. The placement of these complex transitions and the expansion of the level 
scheme were challenging as a result. However, due to the large resolving power and efficiency of Gammasphere 
for high-multiplicity $\gamma$-ray events, the triple-gating technique ($\gamma^{4}$ format) could be used 
to delineate the band structures of interest, rather than the single-gating 
method ($\gamma^{2}$ format) of Ref.~\cite{Theine-NPA-536-418-92}. 
This higher-order gating method proved beneficial, as illustrated by the sample spectra 
of Fig.~\ref{fig:168Ta_band1_2_spe}. All of the tentative transitions proposed earlier~\cite{Theine-NPA-536-418-92} 
have been confirmed, and, in addition, a large number of new $\gamma$ rays were observed 
and placed in the level scheme. As a result, the two structures 
have now been extended significantly, both to lower and higher spins. 
It was also found that the energies of some transitions at the bottom of band 1 were improperly 
assigned earlier~\cite{Theine-NPA-536-418-92}. These $\gamma$ rays and, hence, the associated states 
have been rearranged here (Fig.~\ref{fig:168Ta_level_scheme}), based on the inspection of coincidence 
relationships ($12^{-}{\rightarrow}10^{-}$: 285 changed to 267 keV, $11^{-}{\rightarrow}10^{-}$: 
121 changed to 104 keV, $10^{-}{\rightarrow}9^{-}$: 104 changed to 121 keV). In addition, the 198-keV 
transition at the bottom of band 1 appears to be much stronger than the neighboring 267-keV 
line, as can be seen clearly in the top panel of Fig.~\ref{fig:168Ta_band1_2_spe}. 
To the best of our knowledge, this is an abnormal situation and does not support the view that the 
198-keV transition belongs to band 1. Perhaps, it could correspond to a transition linking band 1 to 
other unidentified structures in $^{168}$Ta. Therefore, the placement of this transition in band 1 
should be viewed as tentative, and the associated $8^{-}$ level 
is drawn as a dashed line in Fig.~\ref{fig:168Ta_level_scheme}. 

In the previous high-spin study~\cite{Theine-NPA-536-418-92}, none of the $\gamma$ rays connecting 
the high-spin sequences to the low-spin states established in the ${\beta}^{+}$ decay of 
$^{168}$W~\cite{Meissner-ZPA-337-45-90} were observed. Thus, spin and parity information could 
not be deduced directly for the observed bands, and possible $I^{\pi}$ quantum numbers and other intrinsic 
properties were assigned tentatively, based solely on the configurations assumed for the bands. 
Here, an extensive study of the configurations, spins, and parities associated with the 
bands was carried out. As a result, general agreement was found with the configurations 
proposed in Ref.~\cite{Theine-NPA-536-418-92}; $\it{i.e.}$, for both bands 
the parity remains the same, but the spin values were reduced by $1\hbar$, 
compared to the previous assignments~\cite{Theine-NPA-536-418-92}. 
In fact, this change in spin is important since it provides consistency 
in the discussion of signature splitting patterns presented below. 
The relevant details on the assignment of configuration, spin, and parity are in the next section. 
Despite an extensive search, no linking transitions between bands 1 and 2 or any new structure 
connected to the known bands were found. Hence, the relative spins and energies of the two bands 
cannot be determined experimentally. 

\section{\label{sec:discussion}Discussion}

To understand band structures in odd-odd nuclei, it is useful to study the additive 
coupling of quasiproton and quasineutron states in neighboring odd-mass 
nuclei ~\cite{Kreiner-PRC-36-2309-87}. This concept has been successfully employed in the $A$ $\sim$ 160--170 
region, see Refs.~\cite{Drissi-NPA-451-313-86,Theine-NPA-536-418-92,Aguilar-170Ta-unpublish-10}, for example. 
In the previous high-spin study of $^{168}$Ta~\cite{Theine-NPA-536-418-92}, the two observed bands 
were interpreted as structures resulting from the coupling of proton states in $^{167}$Ta 
and neutron states in $^{167}$Hf. More specifically, bands 1 and 2 were proposed to have the configurations 
$\pi{h_{11/2}}{[514]}9/2{\otimes}{\nu}i_{13/2}{[642]}5/2$ and $\pi{d_{5/2}}{[402]}5/2{\otimes}{\nu}i_{13/2}{[642]}5/2$, 
respectively. However, this assignment was largely based on the relative intensity of the two bands, 
which may not always be reliable. 

In the present work, the configuration assignments are discussed by considering information on 
energy staggerings, $B(M1)/B(E2)$ ratios, and rotation-induced alignments. 

As described above, the bands in 
$^{168}$Ta can be understood in terms of coupling the single-quasiparticle structures in $^{167}$Ta and 
in $^{167}$Hf. The band of ${\nu}i_{13/2}$ parentage is the yrast and most strongly populated band in $^{167}$Hf. 
Further, this ${\nu}i_{13/2}$ band is characterized by a large signature splitting (${\sim}240~{\rm keV}$ 
at $\hbar\omega{=}0.175~{\rm MeV}$)~\cite{Cromaz-PRC-59-2406-99,Smith-EPJA-6-37-99} and all signature splittings 
observed in $^{168}$Ta are far smaller. Hence, it is the favored $\alpha=+1/2$ sequence of this 
${\nu}i_{13/2}$ structure that couples to both signature partners of a single-quasiproton structure to 
form the strongly-coupled bands in $^{168}$Ta. In the most recent $^{167}$Ta 
data~\cite{Hartley-PRC-80-R041304-09}, six different single-quasiproton structures were observed. 
The bands of ${\pi}h_{9/2}$ and ${\pi}i_{13/2}$ parentage were found to be decoupled (only a single 
sequence was observed for each band), therefore they are unlikely to be involved in forming the bands 
in $^{168}$Ta. The ${\pi}d_{3/2}$ band possesses a large signature splitting (${\sim}$ 100 keV) 
at low spin. The coupling of the two signature partners of ${\pi}d_{3/2}$ to the $\alpha=+1/2$ signature 
of ${\nu}i_{13/2}$ would result in a structure having a comparably large signature splitting. This is 
not the case for the bands in $^{168}$Ta and, therefore, this quasiproton is ruled out as well. 

\begin{figure*}
\begin{center}
\includegraphics[angle=270,width=1.4\columnwidth]{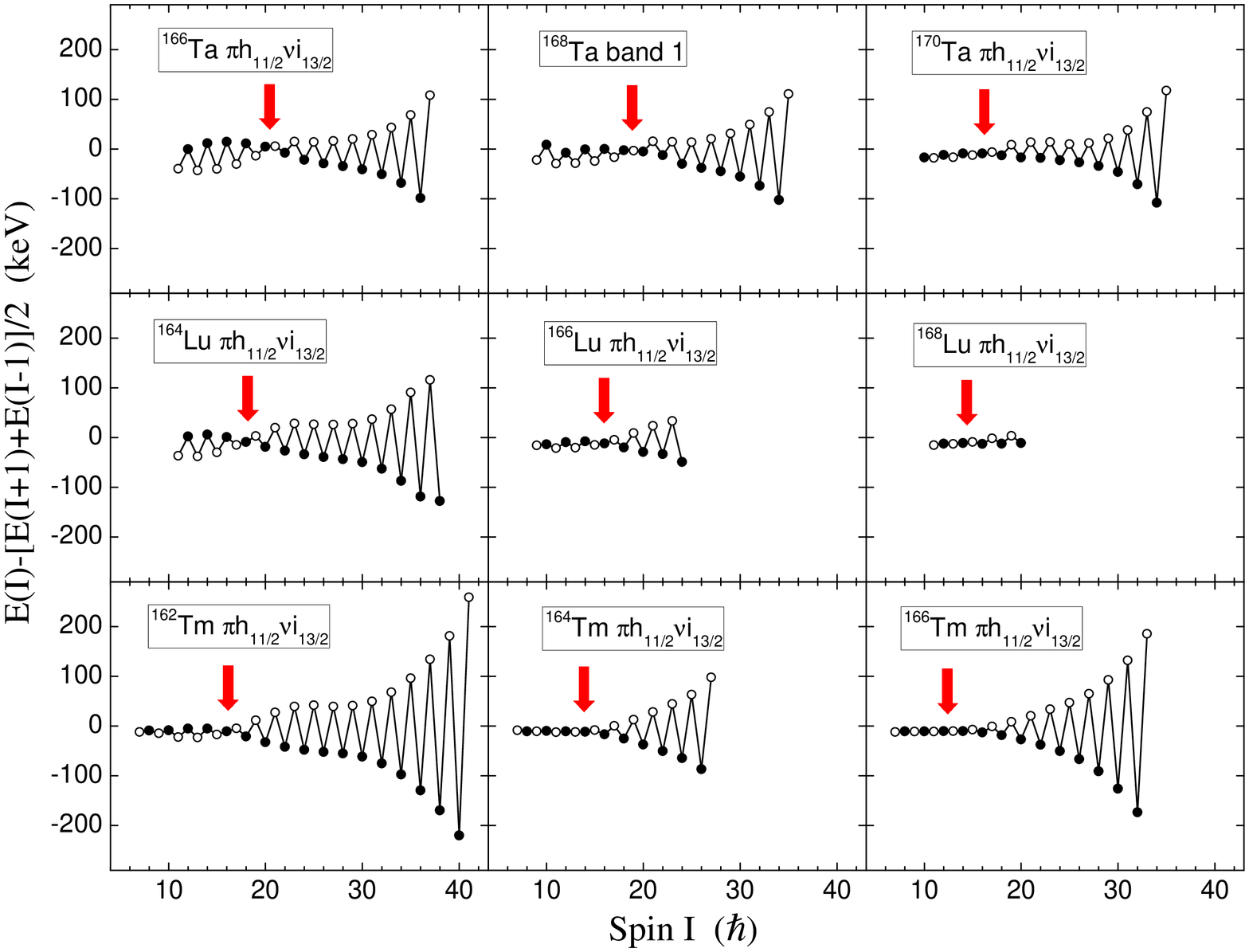}
\caption{(Color on-line) The energy staggerings $E(I)-[E(I+1)+E(I-1)]/2$ as a function of spin $I$ 
for the available $\pi{h_{11/2}}{[514]}9/2{\otimes}{\nu}i_{13/2}{[642]}5/2$ bands in the odd-odd nuclei neighboring 
$^{168}$Ta and band 1 in $^{168}$Ta. The $\alpha=0$ and $\alpha=1$ states are denoted 
by filled and open symbols, respectively. Inversion points are marked by red arrows. 
See text for details.\label{fig:systmt_enerStag_h11-2_i13-2}}
\end{center}
\end{figure*}

\begin{figure*}
\begin{center}
\includegraphics[angle=270,width=1.4\columnwidth]{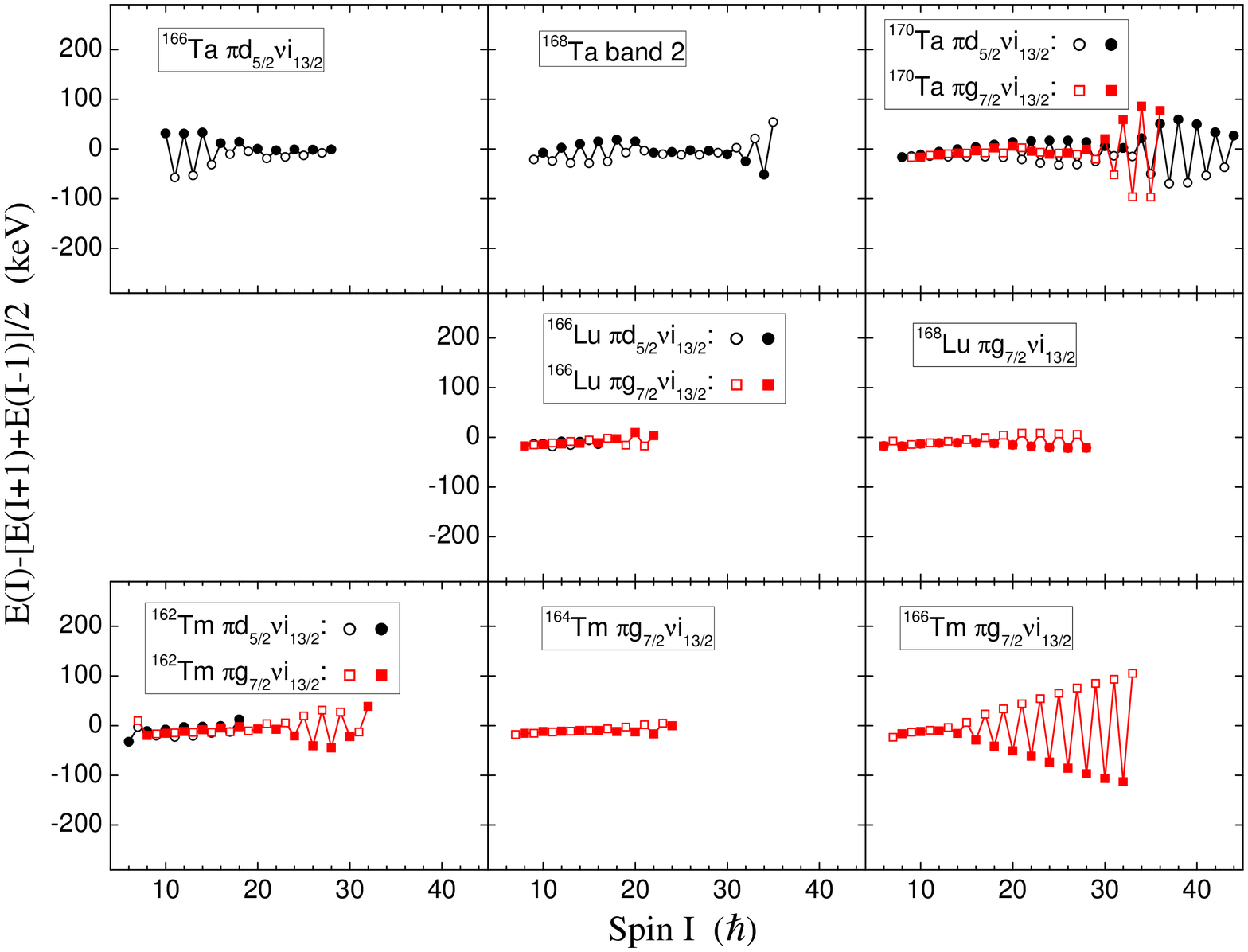}
\caption{(Color on-line) The energy staggerings $E(I)-[E(I+1)+E(I-1)]/2$ as a function of spin $I$ 
for the available $\pi{d_{5/2}}{[402]}5/2{\otimes}{\nu}i_{13/2}{[642]}5/2$ and 
$\pi{g_{7/2}}{[404]}7/2{\otimes}{\nu}i_{13/2}{[642]}5/2$ bands in the odd-odd nuclei neighboring $^{168}$Ta 
and band 2 in $^{168}$Ta. The $\alpha=0$ and $\alpha=1$ states are denoted by filled and open 
symbols, respectively. See text for details.\label{fig:systmt_enerStag_d5-2-g7-2_i13-2}}
\end{center}
\end{figure*}

The energy staggering between signature partners of a rotational band can be studied by comparing the energy of 
a given level with the average of the energies of the signature-partner levels with one unit of spin change 
(higher and lower). The energy staggerings for the two bands of $^{168}$Ta are compared in 
Figs.~\ref{fig:systmt_enerStag_h11-2_i13-2} and \ref{fig:systmt_enerStag_d5-2-g7-2_i13-2} with the 
relevant structures in the neighboring odd-odd nuclei. 
In Fig.~\ref{fig:systmt_enerStag_h11-2_i13-2}, a sharp signature inversion is observed 
at spin $\sim$ 19$\hbar$ in band 1 of $^{168}$Ta. This phenomenon is understood as 
one of the characteristics of the $\pi{h}_{11/2}\nu{i}_{13/2}$ 
structures in odd-odd nuclei in this mass region~\cite{Liu-PRC-52-2514-95}. 
In Fig.~\ref{fig:systmt_enerStag_h11-2_i13-2}, the systematic trends of signature inversion 
for this configuration~\cite{Liu-PRC-52-2514-95} are given for the odd-odd Ta, Lu, and Tm nuclei 
with $N=93$, 95, and 97. In a chain of isotopes, the inversion point moves to lower spin 
with increasing neutron number, and, in a chain of isotones, the inversion point shifts to higher spin with 
increasing proton number. These systematic observations are consistent with the $\pi{h}_{11/2}\nu{i}_{13/2}$ 
configuration assignment to band 1 in $^{168}$Ta. Additional supporting arguments are discussed below. 

For band 2 in $^{168}$Ta, it is more difficult to assign a firm configuration based on the energy staggering 
information. In the energy staggering plot (Fig.~\ref{fig:systmt_enerStag_d5-2-g7-2_i13-2}), 
the $\alpha=1$ sequence of band 2 is lower in energy than its partner ($\alpha=0$) up to spin $\sim$ 
21$\hbar$. By investigating the data available for all $\pi{g}_{7/2}{\nu}i_{13/2}$ and 
$\pi{d}_{5/2}{\nu}i_{13/2}$ bands in the neighboring odd-odd nuclei, band 2 in $^{168}$Ta was found 
to be more consistent with a $\pi{d}_{5/2}\nu{i}_{13/2}$ configuration, based on the observed 
energy staggering pattern at low spin; $\it{i.e.}$, a larger energy difference between the signature partners (as found in band 2) 
is associated with the $\pi{d}_{5/2}\nu{i}_{13/2}$ rather than the $\pi{g}_{7/2}{\nu}i_{13/2}$ configuration. 

It should be noticed that, at the highest spins ($I{\ge}30$), it is the $\alpha=0$ signature 
that is favored in band 2 of $^{168}$Ta. This signature inversion is not found in any other example 
for the $\pi{d_{5/2}}{\nu}i_{13/2}$ configurations in Fig.~\ref{fig:systmt_enerStag_d5-2-g7-2_i13-2}. 
This inversion may result from complex band crossings at higher spin (see below). However, 
similar signature inversions have also been observed at lower spin in $\pi{g}_{7/2}{\nu}i_{13/2}$ 
bands in the odd-odd neighbors, for example, $^{162}$Tm. This indicates that the alternative assignment, 
$\pi{g}_{7/2}{\nu}i_{13/2}$, for band 2 in $^{168}$Ta cannot be ruled out based on the present data. 
Such an observation merits further investigation, especially once higher-spin data in the neighboring 
odd-odd nuclei become available. 

\begin{figure}
\begin{center}
\includegraphics[angle=0,width=0.9\columnwidth]{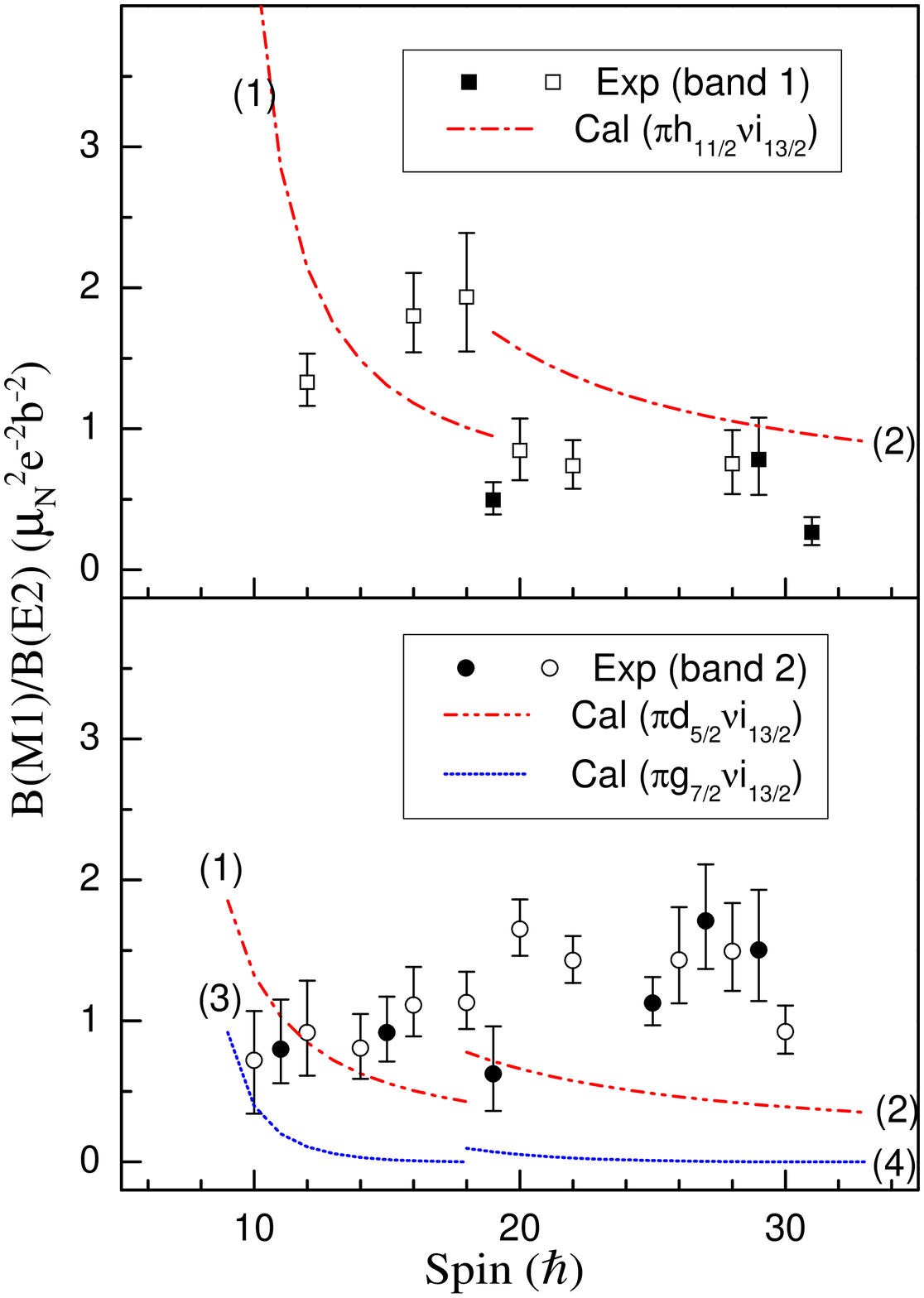}
\caption{(Color on-line) Measured $B(M1)/B(E2)$ ratios as a function of spin for bands 1 (top panel) 
and 2 (bottom panel) in $^{168}$Ta. Filled (open) symbols denote the signatures $\alpha=1(0)$. Theoretical calculations 
are displayed as lines with the following configurations in top panel: (1) $\pi{[514]}9/2{\otimes}{\nu}{[642]}5/2$, 
(2) $\pi{[514]}9/2{\otimes}{\nu}{[642]}5/2BC$; and in bottom panel: (1) $\pi{[402]}5/2{\otimes}{\nu}{[642]}5/2$, 
(2) $\pi{[402]}5/2{\otimes}{\nu}{[642]}5/2BC$, (3) $\pi{[404]}7/2{\otimes}{\nu}{[642]}5/2$, 
(4) $\pi{[404]}7/2{\otimes}{\nu}{[642]}5/2BC$. See Table~\ref{tab:orbt_labl_convt} (and Fig.~\ref{fig:168Ta_neutron_routhian}) 
for the definition of $BC$. See text for details.\label{fig:168Ta_bands1-2_branch_ratio}}
\end{center}
\end{figure}

In order to investigate the configuration assignments for bands 1 and 2 further, $B(M1)/B(E2)$ ratios 
were extracted from the measured $\gamma$-ray intensities. In Fig.~\ref{fig:168Ta_bands1-2_branch_ratio}, 
these ratios are compared with the calculated values for specific configurations. 
The geometric model of D{\"o}nau~\cite{Donau-NPA-471-469-87} was used. Calculations of the 
$B(M1)$ values assumed $g_{R}=Z/A$ and the $g_{\Omega}$ values of Table II in Ref.~\cite{Hartley-PRC-72-064325-05}. 
The standard rotational form of the $B(E2)$ strength~\cite{Bohr-75-book} was employed, assuming a quadrupole moment 
of $6.0~{\rm eb}$ deduced from the predicted deformation of $\beta_{2}=0.24$~\cite{Nazarewicz-NPA-512-61-90}. 
Data for some ratios are missing because the associated inband and/or interband transitions are contaminated 
by other coincident $\gamma$ rays. In the top panel, for band 1, there is good agreement 
between the data and the calculations for the $\pi{h_{11/2}}{\nu}i_{13/2}$ configuration. 
The calculations with the configurations $\pi{g_{7/2}}{\nu}i_{13/2}$ and $\pi{d_{5/2}}{\nu}i_{13/2}$ 
do not reproduce the measured $B(M1)/B(E2)$ ratios in band 2 as well (see bottom panel 
in Fig.~\ref{fig:168Ta_bands1-2_branch_ratio}). However, overall the $\pi{d_{5/2}}{\nu}i_{13/2}$ calculations 
are a closer match to the data. 

From the above analyses, it is concluded that the configuration assignments proposed 
previously~\cite{Theine-NPA-536-418-92}, $\it{i.e.}$, $\pi{h_{11/2}}{[514]}9/2{\otimes}{\nu}i_{13/2}{[642]}5/2$ 
for band 1 and $\pi{d_{5/2}}{[402]}5/2{\otimes}{\nu}i_{13/2}{[642]}5/2$ for band 2, are 
most likely correct. 

\begin{figure}
\begin{center}
\includegraphics[angle=270,width=1.0\columnwidth]{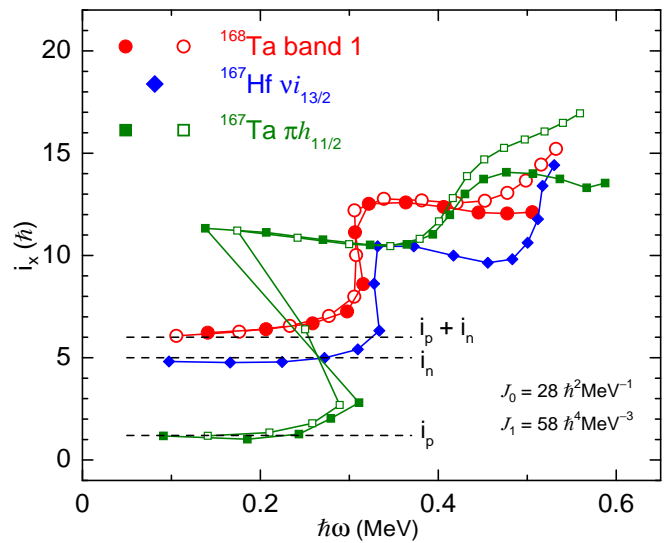}
\caption{(Color on-line) Experimental alignments for band 1 in $^{168}$Ta, the ${\nu}i_{13/2}$ band in $^{167}$Hf, 
and the ${\pi}h_{11/2}$ band in $^{167}$Ta as a function of the rotational frequency. A reference 
with the common $J_{0}$ and $J_{1}$ Harris parameters indicated in the figure was subtracted for each band. 
Filled (open) symbols denote the signatures $\alpha={+}\frac{1}{2}$(${-}\frac{1}{2}$) or 1(0). 
Tentative transitions are excluded from the figure. See text for details.\label{fig:168Ta_band1_alignment}}
\end{center}
\end{figure}

\begin{figure}
\begin{center}
\includegraphics[angle=270,width=1.0\columnwidth]{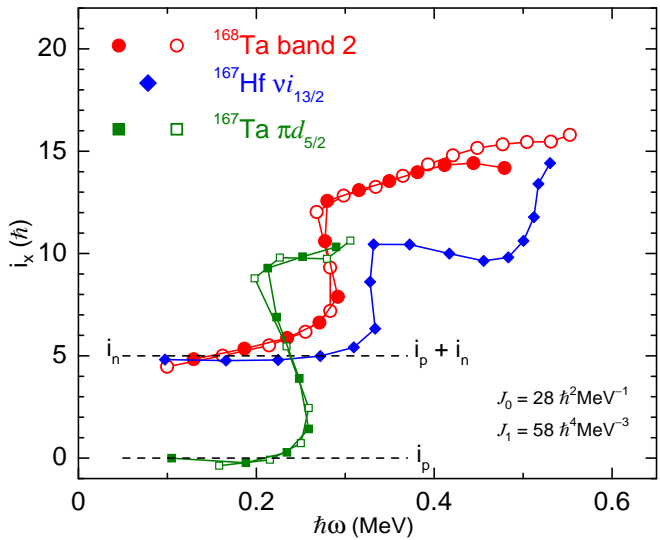}
\caption{(Color on-line) Experimental alignments for band 2 in $^{168}$Ta, the ${\nu}i_{13/2}$ band in $^{167}$Hf, 
and the ${\pi}d_{5/2}$ band in $^{167}$Ta as a function of the rotational frequency. A reference 
with the common $J_{0}$ and $J_{1}$ Harris parameters given in the figure was subtracted for each band. 
Filled (open) symbols denote the signatures $\alpha={+}\frac{1}{2}$(${-}\frac{1}{2}$) or 1(0). 
Tentative transitions are excluded from the figure. See text for details.\label{fig:168Ta_band2_alignment}}
\end{center}
\end{figure}

In the literature~\cite{Theine-NPA-536-418-92}, 
the spins of the levels in $^{168}$Ta were proposed based on an analysis of the additivity of alignments. 
According to the Gallagher-Moszkowski rule~\cite{Gallagher-PR-111-1282-58} the lowest states 
correspond to the parallel case of a quasiparticle coupling; therefore, $K=7$ and $K=5$ 
($K=\Omega_p+\Omega_n$) were assigned to bands 1 ($\pi{h_{11/2}}{[514]}9/2{\otimes}{\nu}i_{13/2}{[642]}5/2$) 
and 2 ($\pi{d_{5/2}}{[402]}5/2{\otimes}{\nu}i_{13/2}{[642]}5/2$), respectively. In Ref.~\cite{Theine-NPA-536-418-92}, 
when plotting the alignments for the bands, different Harris parameters were used for each 
nucleus. However, in the present work, a common set of Harris parameters 
from Ref.~\cite{Hartley-PRC-74-054314-06} (${J}_0=28~{\hbar}^2{\rm MeV}^{-1}$ and 
${J}_1=58~{\hbar}^4{\rm MeV}^{-3}$) was adopted such that the ground-state band in the 
neighboring even-even $^{168}$Hf nucleus has nearly zero initial alignment and a 
constant alignment above the first crossing. The extracted aligned spins for bands of interest 
in $^{167}$Ta, $^{167}$Hf, and $^{168}$Ta are plotted in Figs.~\ref{fig:168Ta_band1_alignment} 
and \ref{fig:168Ta_band2_alignment}. As in Ref.~\cite{Theine-NPA-536-418-92}, 
the spin was assigned such that the alignment additivity rule is realized for each band; $\it{i.e.}$, 
$i_{pn}$($^{168}$Ta)$=i_p$($^{167}$Ta)$+i_n$($^{167}$Hf). It was found that the best quantitative 
agreement is reached if $I=8\hbar$ is assigned to the lowest state in each of the bands 1 and 2, 
as seen in Figs.~\ref{fig:168Ta_band1_alignment} and \ref{fig:168Ta_band2_alignment}. 
In other words, the spins of states in both bands have been reduced by $1\hbar$ relative to 
the assignments proposed in Ref.~\cite{Theine-NPA-536-418-92} (see the level scheme in 
Fig.~\ref{fig:168Ta_level_scheme}). These updated spin assignments for the two bands in $^{168}$Ta 
can be viewed as further evidence for the validity of the energy staggering patterns emerging 
in Figs.~\ref{fig:systmt_enerStag_h11-2_i13-2} and \ref{fig:systmt_enerStag_d5-2-g7-2_i13-2} 
(see discussion above). Spins were then proposed for all new, higher-lying levels observed in 
the present work by assuming that the rotational behavior persists throughout the bands. 
The parity of the states has not been measured directly. Instead, it is based on the proposed 
configurations associated with the bands. Consequently, bands 1 and 2 are assumed to 
have negative and positive parity, respectively, again in agreement with Ref.~\cite{Theine-NPA-536-418-92}. 

\begin{figure}
\begin{center}
\includegraphics[angle=90,width=1.0\columnwidth]{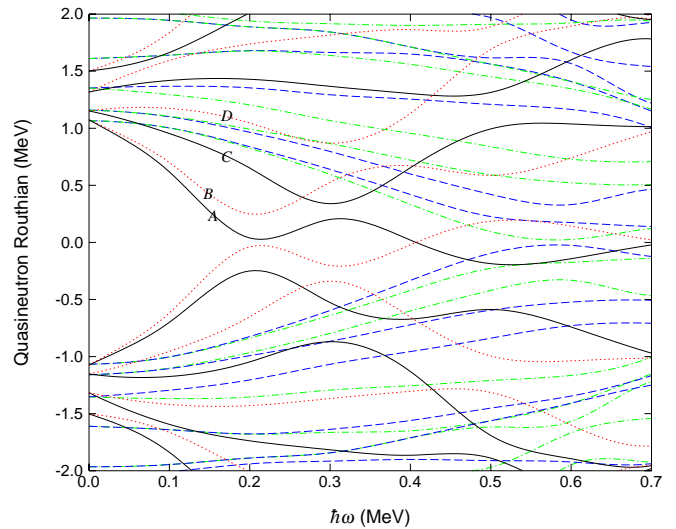}
\caption{(Color on-line) Quasineutron Routhians for $^{168}$Ta calculated in the CSM with deformation parameters 
of $\beta_2=0.24$, $\beta_4=0$, and $\gamma=0$. The four line types denote four combinations of (parity, signature), 
solid: ($+$,$+1/2$); dot: ($+$,$-1/2$); dash-dot: ($-$,$+1/2$); dash: ($-$,$-1/2$). 
See text for details.\label{fig:168Ta_neutron_routhian}}
\end{center}
\end{figure}

\begin{figure}
\begin{center}
\includegraphics[angle=90,width=1.0\columnwidth]{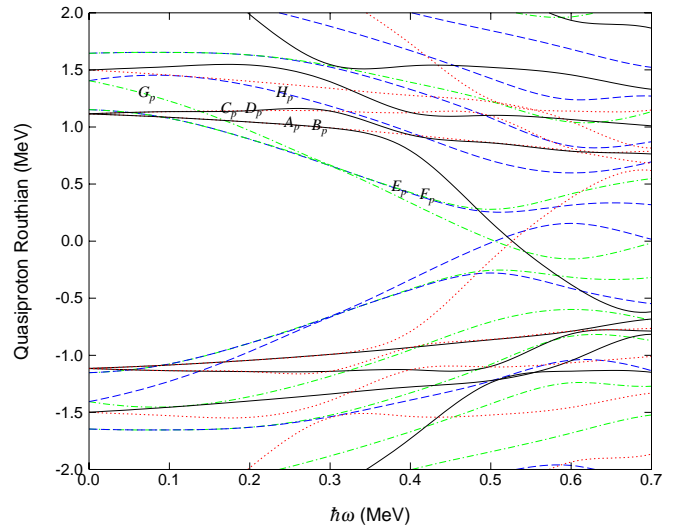}
\caption{(Color on-line) Quasiproton Routhians for $^{168}$Ta calculated in the CSM with deformation parameters 
of $\beta_2=0.24$, $\beta_4=0$, and $\gamma=0$. The convention for drawing the lines is same as that used in 
Fig.~\ref{fig:168Ta_neutron_routhian}. See text for details.\label{fig:168Ta_proton_routhian}}
\end{center}
\end{figure}

In order to interpret the variations in alignment properties of the rotational sequences as a function of rotational frequency, 
Cranked Shell Model (CSM) calculations have been performed for both quasiprotons and quasineutrons in $^{168}$Ta. 
In the calculations, deformation parameters of $\beta_2=0.24$, $\beta_4=0$, and $\gamma=0$ were adopted, 
as suggested in Ref.~\cite{Nazarewicz-NPA-512-61-90}. The pairing energies were estimated 
from the mass differences of neighboring nuclei, using the five-point 
fit defined in Ref.~\cite{Moller-NPA-536-20-92}. The resulting quasineutron and quasiproton Routhians 
are displayed in Figs.~\ref{fig:168Ta_neutron_routhian} and \ref{fig:168Ta_proton_routhian}, respectively. 
Quasiparticle orbitals close to the Fermi surface in these figures have been labeled using the Stockholm 
convention~\cite{Bengtsson-ADNDT-35-15-86}, and are summarized for $^{168}$Ta in Table~\ref{tab:orbt_labl_convt}. 

\begin{table}
\caption{The convention used to label the quasiparticle orbitals in the Routhian 
figures and in the discussion in the text.\label{tab:orbt_labl_convt}}
\begin{ruledtabular}
\begin{tabular}{ccc}
Shell-model state & Nilsson label & Label used ($\alpha=+\frac{1}{2}$, $-\frac{1}{2}$)\\ \hline
\\
$\nu{i}_{13/2}$ & $\nu[642]5/2$ & $A$, $B$\\
$\nu{i}_{13/2}$ & $\nu[633]7/2$ & $C$, $D$\\
$\pi{d}_{5/2}$ & $\pi[402]5/2$ & $A_p$, $B_p$\\
$\pi{g}_{7/2}$ & $\pi[404]7/2$ & $C_p$, $D_p$\\
$\pi{h}_{11/2}$ & $\pi[514]9/2$ & $E_p$, $F_p$\\
$\pi{h}_{9/2}$ & $\pi[541]1/2$ & $G_p$, $H_p$\\
\end{tabular}
\end{ruledtabular}
\end{table}

As seen in the alignment plot (Fig.~\ref{fig:168Ta_band1_alignment}), the first discontinuity (upbending or backbending) 
for both signatures of band 1 in $^{168}$Ta occurs at a frequency of $\hbar\omega{\sim}0.31~{\rm MeV}$. 
Based on the frequency and the gain in angular momentum ($\sim{5}\hbar$), this alignment can be understood as 
resulting from the $BC$ neutron crossing, which is predicted to occur at $\hbar\omega{\sim}0.31~{\rm MeV}$ in the CSM 
calculations with a calculated alignment of $5-6\hbar$, see Fig.~\ref{fig:168Ta_neutron_routhian}. 
The $\nu{i}_{13/2}$ band observed in $^{167}$Hf exhibits a similar first alignment (Fig.~\ref{fig:168Ta_band1_alignment}) 
and is consistent with this interpretation. A similar alignment (with the same gain, but at slightly lower frequency) is 
observed in band 2 (Fig.~\ref{fig:168Ta_band2_alignment}). It can also be interpreted as associated with 
the $BC$ neutron crossing. The fact that the backbendings observed at $\hbar\omega{\sim}0.24~{\rm MeV}$ 
($AB$ crossing) in both the $\pi{h}_{11/2}$ (Fig.~\ref{fig:168Ta_band1_alignment}) and $\pi{d}_{5/2}$ 
(Fig.~\ref{fig:168Ta_band2_alignment}) bands of $^{167}$Ta are Pauli blocked in the two bands of $^{168}$Ta, 
due to the occupation of the single neutron $A$ ($i_{13/2}$) orbital, is consistent with the interpretation. 
The small difference in the first band-crossing frequencies between the two bands in $^{168}$Ta 
may possibly reflect a slight difference in deformation or a small change in the value of the 
proton-neutron residual interaction between the odd particles. 

Due to the significant extension of the rotational sequences achieved in the present analysis, a second band crossing 
is revealed for the first time in $^{168}$Ta. As can be seen in Fig.~\ref{fig:168Ta_band1_alignment}, the two 
signature partners of band 1 split above $\hbar\omega{\sim}0.4~{\rm MeV}$. In the $\alpha=0$ sequence, 
the second discontinuity starts at $\hbar\omega{\sim}0.47~{\rm MeV}$, while the $\alpha=1$ partner 
does not display any sign of a second alignment up to the highest frequency ($\hbar\omega{\sim}0.51~{\rm MeV}$) observed. 
Such a difference between signatures also takes place at a similar frequency for the $\pi{h}_{11/2}$ band in $^{167}$Ta 
(Fig.~\ref{fig:168Ta_band1_alignment}). This occurs in $^{167}$Ta after a common alignment gain of $\sim$ $4\hbar$ between 
0.35 and 0.45 MeV in rotational frequency. The latter alignment is also present in the yrast band of 
neighboring even-even nuclei, where it is interpreted as the $CD$ neutron crossing. Such a crossing ($CD$) is 
Pauli blocked in the $\nu{i_{13/2}}$ band of $^{167}$Hf ($ABC$ in the frequency range discussed) 
as well as in band 1 of $^{168}$Ta ($ABCE_{p}$ and $ABCF_{p}$ in the frequency range discussed). 
For the splitting in alignment of the two signatures in band 1 of $^{168}$Ta at high frequency 
($\hbar\omega{\ge}0.4~{\rm MeV}$), inspection of the calculated quasineutron Routhians 
(Fig.~\ref{fig:168Ta_neutron_routhian}) reveals that there are no suitable neutron crossings close to this 
frequency ($\hbar\omega{\sim}0.5~{\rm MeV}$) that could account for the observation. 
This suggests that the second alignment observed in the $\alpha=0$ signature of band 1 in $^{168}$Ta may 
arise from a proton crossing. 

Quasiproton Routhians calculated using the CSM (Fig.~\ref{fig:168Ta_proton_routhian}) exhibit a complex set 
of crossings at frequencies near $\hbar\omega{\sim}0.5~{\rm MeV}$, due in part to the presence of a favored 
$h_{9/2}$ orbital ($G_p$). The crossings $E_{p}F_p$, $G_{p}H_p$, $F_{p}G_p$, and $E_{p}H_p$ are all 
calculated to occur approximately at the frequency of interest. In fact, the $\pi{h}_{11/2}$ and $\pi{h}_{9/2}$ 
orbitals are well known to be strongly mixed at high rotational 
frequencies~\cite{Cromaz-PRC-59-2406-99,Smith-EPJA-6-37-99,Hartley-PRC-72-064325-05,Hartley-PRC-74-054314-06}. 
It is tentatively proposed that the second discontinuity in the aligned spin observed in band 1 is caused 
by the mixed $E_p{H}_{p}$ proton crossing. Such mixed crossings have previously been discussed in the 
osmium-iridium region~\cite{Marshalek-PRL-55-370-85}. Furthermore, the same orbitals have been proposed 
to explain similar high-frequency alignments in neighboring $^{169}$Hf~\cite{Gao-PRC-44-1380-91}. 

The single quasiproton $F_{p}$ in the even-spin sequence ($\alpha=0$) of band 1 in $^{168}$Ta would block the 
$E_p{F_p}$ crossing, but not the $E_p{H}_{p}$ one. Thus, in the context of the present interpretation of a mixed 
$E_p{H}_{p}$ proton crossing, it is expected that, at a frequency of ${\sim}0.5~{\rm MeV}$, the second alignment occurs in 
the sequence with signature $\alpha=0$, while it is not present in the $\alpha=1$ partner. 
This is consistent with the observations in band 1 of $^{168}$Ta (Fig.~\ref{fig:168Ta_band1_alignment}). 
However, the CSM calculations do not reproduce some of the associated proton orbitals, 
as can be seen in Fig.~\ref{fig:168Ta_proton_routhian}. The $G_p$ orbital is positioned low in energy, 
and becomes yrast at $\hbar\omega{>}0.3~{\rm MeV}$. This results in the prediction that the $G_{p}H_p$ and $F_{p}G_p$ 
crossings should also occur at frequencies between 0.5 and 0.6 MeV, similar to the $E_{p}H_p$ one. 
If the $G_p{H}_{p}$ or $F_p{G}_{p}$ crossing or both were involved, the two signatures in band 1 in $^{168}$Ta 
would encounter the second alignment in a similar way, in contradiction with the experimental observations. 
The difficulty of the CSM in reproducing proton crossings at high frequencies for nuclei in this region is 
well documented. Such discrepancies may come, for example, from a change of deformation with increasing 
frequency and/or higher-order pairing terms, which are not contained in the CSM~\cite{Hartley-PRC-72-064325-05,Jensen-NPA-695-3-01}. 

For band 2 in $^{168}$Ta, the existence of the $d_{5/2}$ ($A_p$, $B_p$) quasiproton in its configuration would not 
prevent the occurrence of the mixed $E_p{H}_{p}$ (or $E_p{F_p}$) proton crossing in both signatures. 
However, as can be seen in Fig.~\ref{fig:168Ta_band2_alignment}, the $E_p{H}_{p}$ alignment is delayed up to 
the highest frequency observed in the ${\alpha}=0$ sequence (${\le}0.56~{\rm MeV}$) and the ${\alpha}=1$ signature (${\le}0.48~{\rm MeV}$). 
It is possible that the $E_p{F}_{p}$ crossing occurs with a large interaction strength and results in a smooth 
gradual alignment for both signatures of band 2, which would block the subsequent $E_{p}H_p$ and $F_p{G}_{p}$ crossings. 

From the detailed discussion above, it is clear that the understanding of the proton alignments 
involving $h_{11/2}$ and $h_{9/2}$ orbitals that occur in this mass region is far from complete and 
that further investigation is necessary. 

\section{\label{sec:summary}Summary and Conclusions}
The two coupled bands previously observed in $^{168}$Ta have been extended 
significantly to higher spin, from $\sim{20}\hbar$ to above $40\hbar$. 
Quasiparticle configurations were assigned to the bands based on the observed signature splittings, 
$B(M1)/B(E2)$ ratios, and rotation-induced alignments. A $\pi{h_{11/2}}{\nu}i_{13/2}$ configuration 
has been assigned to band 1, while band 2 is proposed to be associated with the $\pi{d_{5/2}}{\nu}i_{13/2}$ 
configuration. Revised spin values for the two bands have been proposed, based on the concepts of additivity and 
signature staggering. A second discontinuity in aligned spin has been observed for the first time in this nucleus 
and interpreted as a quasiproton band crossing. The observed alignment behaviors are found to be consistent with 
similar observations in the neighboring $^{167,169,171}$Ta (odd-$Z$) and $^{167}$Hf (odd-$N$) nuclei. 
In the CSM calculations, a complex pattern of alignments involving $h_{11/2}$ and $h_{9/2}$ 
proton orbitals is predicted at high rotational frequencies. A fully consistent picture has yet to 
emerge in order to explain the high-frequency alignments observed in $^{168}$Ta and other 
nuclei in this mass region. A dedicated experiment optimized to populate $^{168}$Ta would 
likely reveal additional band structures, such as those associated with the $\pi{h_{9/2}}{\nu}i_{13/2}$ 
configuration which is expected to be yrast at the highest spins. Such a measurement could help resolve 
the alignment puzzles at high frequency discussed in the present work. 

\section*{ACKNOWLEDGMENTS}
The authors gratefully acknowledge J. P. Greene for the preparation of the targets, and the ATLAS operations 
staff for providing the high-quality beam. We also thank D. C. Radford and H. Q. Jin for their 
software support. 

This work has been supported in part by the U.S. National Science Foundation under grants 
No. PHY-0554762 (USNA), PHY-0456463 (FSU), and PHY-0754674 (UND), the U.S. Department of Energy, 
Office of Nuclear Physics, under contracts No. DE-AC02-06CH11357 (ANL), DE-FG02-94ER40848 (UML), and DE-FG02-96ER40983 (UTK), 
and the State of Florida.

\end{document}